\documentclass[preprint,12pt]{elsarticle}

\usepackage{amssymb}

\usepackage{amsmath}

\begin{document}

\begin{frontmatter}



\title{Supersymmetry restoration in lattice formulations of 2D
$\mathcal{N}=(2,2)$ WZ model based on the Nicolai map}


\author{Daisuke Kadoh}
\ead{kadoh@riken.jp}

\author{Hiroshi Suzuki}
\ead{hsuzuki@riken.jp}

\address{Theoretical Physics Laboratory, RIKEN, Wako 2-1, Saitama 351-0198,
Japan}

\begin{abstract}
For lattice formulations of the two-dimensional $\mathcal{N}=(2,2)$
Wess--Zumino (2D $\mathcal{N}=(2,2)$ WZ) model on the basis of the Nicolai map,
we show that supersymmetry (SUSY) and other symmetries are restored in the
continuum limit without fine tuning, to all orders in perturbation theory. This
provides a theoretical basis for use of these lattice formulations for
computation of correlation functions.
\end{abstract}



\end{frontmatter}


\section{Introduction}
It is believed that at long distance, 2D $\mathcal{N}=(2,2)$ WZ model with a
quasi-\hspace{0pt}homogeneous superpotential\footnote{\label{fn:one}%
A polynomial $W(\phi)$ of variables $\phi_I$ ($I=1$, $2$, \dots, $N$) is called
quasi-homogeneous, when there exist some weights~$\omega_I$ such that
$W(\phi_I\to\Lambda^{\omega_I}\phi_I)=\Lambda W(\phi)$.} provides a
Landau--Ginzburg description of $\mathcal{N}=(2,2)$ superconformal field
theories (SCFT)~\cite{Kastor:1988ef,Vafa:1988uu,Lerche:1989uy,Howe:1989qr,%
Cecotti:1989jc,Howe:1989az,Cecotti:1989gv,Cecotti:1990kz,Witten:1993jg}. See
\S14.4 of~Ref.~\cite{Hori:2003ic} for a review. Although this expectation has
been tested in various ways, it is very difficult to confirm this WZ/SCFT
correspondence directly by comparing general correlation functions in both
theories; 2D WZ model is strongly coupled in low energies and for such a
comparison, one needs a certain powerful tool which enables nonperturbative
calculation.

In a recent paper~\cite{Kawai:2010yj}, Kawai and~Kikukawa reconsidered this
problem and they computed some correlation functions in 2D WZ model by
numerical simulation of a lattice formulation developed
in~Ref.~\cite{Kikukawa:2002as}. They considered the WZ model with a cubic
superpotential $W(\phi)=\lambda\phi^3/3$, which, according to the conjectured
correspondence, should provide a Landau--Ginzburg description of the
$A_2$~model. The central charge of the $A_2$~model is~$c=1$ (the gaussian
model) and a (unique) chiral primary field in the NS sector, $\Phi_{0,0}$, which
should be given by the scalar field of the WZ model in the infrared, has
conformal dimensions $(h,\overline h)=(1/6,1/6)$. Finite-size scalings of
scalar two-point functions observed in~Ref.~\cite{Kawai:2010yj} are remarkably
consistent with the above expectation. Ref.~\cite{Kawai:2010yj} thus certainly
demonstrated a use for lattice formulations in studying nonperturbative
dynamics of supersymmetric field theory (there exist preceding numerical
simulations of the 2D $\mathcal{N}=(2,2)$ WZ model with a massive cubic
superpotential $W(\phi)=m\phi^2/2+\lambda\phi^3/3$~\cite{Beccaria:1998vi,%
Catterall:2001fr,Giedt:2005ae,Bergner:2007pu,Kastner:2008zc}).

Having observed the success of~Ref.~\cite{Kawai:2010yj}, one is naturally lead
to consider the 2D $\mathcal{N}=(2,2)$ WZ model with more general
(quasi-homogeneous) superpotentials. It would be interesting to generalize the
study of~Ref.~\cite{Kawai:2010yj} to $W(\phi)=\lambda\phi^n/n$ with~$n>3$, for
example, which is thought to correspond to the $A_{n-1}$~model, or to
$W(\phi)=\lambda\phi^n/n+\lambda'\phi\phi^{\prime2}/2$ with~$n\geq3$, where
$\phi$ and~$\phi'$ are independent scalar fields, which should correspond to
the $D_{n+1}$~model.

Before going into such study of physical questions, however, one has to be sure
at least within perturbation theory\footnote{In the context of the
Landau--Ginzburg description of nontrivial SCFT, one is interested in the WZ
model without mass term for which, strictly speaking, 2D perturbation theory is
a formal one due to severe infrared divergences. Thus, it is eventually
desirable to confirm the symmetry restoration in a non-perturbative manner, as
had been done in~Ref.~\cite{Kanamori:2008bk} for the 2D $\mathcal{N}=(2,2)$
supersymmetric Yang--Mills theory.} that symmetries which are broken by lattice
regularization (including SUSY) are restored in the continuum limit without
tuning lattice parameters. Somewhat surprisingly, such an argument for symmetry
restoration in lattice formulations of the 2D $\mathcal{N}=(2,2)$ WZ model is
not found in the literature, except those for the cubic superpotential with a
single supermultiplet: Ref.~\cite{Giedt:2004qs} for a lattice formulation
of~Ref.~\cite{Sakai:1983dg} and Ref.~\cite{Kawai:2010yj} for a formulation
of~Ref.~\cite{Kikukawa:2002as}. In fact, at first glance, it appears that
rather complicated enumeration of possible symmetry breaking operators is
required for an argument for general superpotentials. The purpose of the
present article is to point out that there actually exists a very simple way to
see the symmetry restoration in the continuum limit for lattice
formulations~\cite{Sakai:1983dg,Kikukawa:2002as,Bergner:2007pu} based on the
Nicolai map~\cite{Nicolai:1979nr,Nicolai:1980jc,Cecotti:1981fu,Parisi:1982ud,%
Cecotti:1982ad} for general superpotentials. We can show that SUSY and other
symmetries are restored in the continuum limit without fine tuning to all
orders of perturbation theory.\footnote{There also exists a valid lattice
formulation of the 2D $\mathcal{N}=(2,2)$ WZ model on the basis of the SLAC 
derivative in which SUSY and other symmetries are
manifest~\cite{Bartels:1983wm,Kadoh:2009sp}.}

\section{Lattice formulations based on the Nicolai map}

Lattice formulations of 2D $\mathcal{N}=(2,2)$ WZ model based on the Nicolai
map~\cite{Sakai:1983dg,Kikukawa:2002as,Bergner:2007pu} can be succinctly
expressed in the following form ($a$ denotes the lattice spacing):
\begin{align}
   S_{\text{2DWZ}}^{\text{LAT}}
   &=Q
   a^2\sum_x\left[
   -\psi_{I-}G_I-\psi_{I+}\eta_I(\phi,\phi^*)-\psi_{I-}\eta_I^*(\phi,\phi^*)
   \right]
\notag\\
   &=a^2\sum_x\Biggl[
   -G_I^*G_I-G_I\eta_I(\phi,\phi^*)-G_I^*\eta_I^*(\phi,\phi^*)
\notag\\
   &\qquad\qquad\qquad\qquad\qquad
   -(\psi_{I+},\psi_{I-})
   \begin{pmatrix}
   \frac{\partial\eta_I}{\partial\phi_J}
   &\frac{\partial\eta_I}{\partial\phi_J^*}\\
   \frac{\partial\eta_I^*}{\partial\phi_J}
   &\frac{\partial\eta_I^*}{\partial\phi_J^*}
   \end{pmatrix}
   \begin{pmatrix}
   \overline\psi_{J-}\\\overline\psi_{J+}
   \end{pmatrix}\Biggr],
\label{eq:one}
\end{align}
where $(\phi_I^{(*)},\psi_{\pm I},\overline\psi_{\mp I},G_I^{(*)})$ ($I=1$, $2$,
\dots, $N$) denotes a supermultiplet and the summation over repeated ``flavor''
indices $I$, $J$, \dots\ is understood; the superscript in the form~$x^{(*)}$
implies either $x$ or~$x^*$ throughout this article. $Q$~is one particular
spinor component of the $\mathcal{N}=(2,2)$ super transformation\footnote{The
explicit form of the $\mathcal{N}=(2,2)$ super transformation can be found, for
example, in Appendix~A of~Ref.~\cite{Sugino:2008yp}. Spinor components in the
present article and those in~Ref.~\cite{Sugino:2008yp} are related by:
$\psi_+=\psi_R$, $\psi_-=\overline\psi_L$, $\overline\psi_-=\psi_L$
and~$\overline\psi_+=\overline\psi_R$.} and its explicit form is given by
\begin{align}
   &Q\phi_I=-\overline\psi_{I-},&& Q\overline\psi_{I-}=0,
\notag\\
   &Q\phi_I^*=-\overline\psi_{I+},&& Q\overline\psi_{I+}=0,
\notag\\
   &Q\psi_{I+}=G_I,&& QG_I=0,
\notag\\
   &Q\psi_{I-}=G_I^*,&& QG_I^*=0.
\label{eq:two}
\end{align}
Since this fermionic transformation is nilpotent, $Q^2=0$, the lattice
action~\eqref{eq:one} is manifestly invariant under this transformation,
$QS_{\text{2DWZ}}^{\text{LAT}}=0$, for any choice of the
functions~$\eta_I(\phi,\phi^*)$. Actually, lattice actions
in~Refs.~\cite{Sakai:1983dg,Kikukawa:2002as,Bergner:2007pu} are actions
obtained after integrating over the auxiliary fields~$G_I$ ($G_I$ is a
``shifted'' auxiliary field and in the continuum theory, it is defined from
the conventional auxiliary field~$F_I$ by
$G_I\equiv F_I+(\partial_0+i\partial_1)\phi_I$). In this article, we instead
use representation~\eqref{eq:one} because with explicit auxiliary fields, the
$Q$~transformation is nilpotent even without using the equation of motion. The
action~\eqref{eq:one} is also invariant under the $U(1)_V$
transformation,\footnote{The \emph{continuum\/} action of the 2D
$\mathcal{N}=(2,2)$ WZ model possesses another $R$-symmetry, a $\mathbb{Z}_2$
symmetry, that is defined by~$\phi_I\leftrightarrow\phi_I^*$,
$\psi_I\leftrightarrow i\sigma_2\overline\psi_I^T$
and~$F_I\leftrightarrow F_I^*$.}
\begin{equation}
   \psi_I\equiv
   \begin{pmatrix}
   \psi_{I+}\\\psi_{I-}
   \end{pmatrix}
   \to e^{-i\alpha}\psi_I,\qquad
   \overline\psi_I\equiv(\overline\psi_{I-},\overline\psi_{I+})
   \to e^{i\alpha}\overline\psi_I.
\label{eq:three}
\end{equation}

Although the $Q$-invariance of~Eq.~\eqref{eq:one} holds for any choice
of~$\eta_I(\phi,\phi^*)$, for the lattice action to have a correct classical
continuum limit, $\eta_I(\phi,\phi^*)$ should become in the classical continuum
limit a combination that specifies the Nicolai map in 2D $\mathcal{N}=(2,2)$ WZ
model, $\eta_I(\phi,\phi^*)\xrightarrow{a\to0}%
\partial W(\phi)/\partial\phi_I-\left(\partial_0-i\partial_1\right)\phi_I^*$.
(The Nicolai map in 2D $\mathcal{N}=(2,2)$ WZ model is the field transformation
from $(\phi,\phi^*)$ to the combination in the right-hand side and its complex
conjugate.) Here, $W(\phi)$ is the superpotential, a holomorphic polynomial of
scalar fields~$\phi_I$,
\begin{equation}
   W(\phi)=\sum_{\{m\}}\frac{\lambda_{\{m\}}}{\prod_{m_I\neq0}m_I}
   \phi_1^{m_1}\phi_2^{m_2}\dotsb\phi_N^{m_N},
\label{eq:four}
\end{equation}
and $\{m\}\equiv\{m_1,m_2,\dotsc,m_N\}$ is a collection of non-negative
integers. In what follows, we assume that field variables are chosen so that
$W(\phi)$ and thus the scalar potential in the WZ model,
$V(\phi,\phi^*)=\sum_I|\partial_IW(\phi)|^2$, do not have any linear tadpole
terms. Note that mass dimensions of the scalar fields~$\phi_I$, the
spinor fields~$\psi_I$ and the auxiliary fields~$G_I$ are $0$, $1/2$ and~$1$,
respectively. As a consequence, all the coupling constants $\lambda_{\{m\}}$
in~Eq.~\eqref{eq:four} have the mass dimension~$1$. Also, as an additional
requirement, the functions $\eta_I(\phi,\phi^*)$ should be chosen such that the
resulting lattice Dirac operator does not have the species doublers.

In the present lattice system~\eqref{eq:one}, the partition function can
(almost) be trivialised as in the continuum theory~\cite{Nicolai:1979nr,%
Nicolai:1980jc,Cecotti:1981fu,Parisi:1982ud,Cecotti:1982ad}, by changing
bosonic integration variables from $(\phi,\phi^*)$ to~$(\eta,\eta^*)$. The
Jacobian associated with this change of variables precisely cancels the
absolute value of the fermion determinant and then the functional integral
becomes (after integrating over the auxiliary fields) gaussian one up to a sign
factor associated with the fermion determinant. This ``almost trivialized''
representation provides a remarkable simulation algorithm that is completely
free from the critical slowing down and a usual difficulty of massless
fermions. See~Refs.~\cite{Beccaria:1998vi,Kawai:2010yj}.

So far, three different choices of $\eta_I(\phi,\phi^*)$ (lattice Nicolai map
function) have been studied. In~Ref.~\cite{Sakai:1983dg}, the authors
adopted (see Refs.~\cite{Elitzur:1982vh,Cecotti:1982ad,Elitzur:1983nj} for
corresponding Hamiltonian formulations)
\begin{equation}
   \eta_I(\phi,\phi^*)=\frac{\partial W(\phi)}{\partial\phi_I}
   -\left(\partial_0^S-i\partial_1^S\right)\phi_I^*
   -\frac{a}{2}\sum_\mu\partial_\mu^*\partial_\mu\phi_I,
\label{eq:five}
\end{equation}
where $\partial_\mu^S\equiv(\partial_\mu^*+\partial_\mu)/2$ and $\partial_\mu$
and~$\partial_\mu^*$ are the forward and backward lattice difference operators,
respectively. This choice of the lattice Nicolai map function leads to (we set
$\gamma_0\equiv\sigma_1$, $\gamma_1\equiv-\sigma_2$
and~$\gamma_5\equiv i\gamma_0\gamma_1=\sigma_3$),
\begin{align}
   S_{\text{2DWZ}}^{\text{LAT}}
   &=a^2\sum_x\biggl[
   -G_I^*G_I-G_I\eta_I(\phi,\phi^*)-G_I^*\eta_I^*(\phi,\phi^*)
\notag\\
   &\qquad\qquad
   +\overline\psi_I\left(D_{\text{w}}
   +\frac{\partial^2W(\phi)}{\partial\phi_I\partial\phi_J}\frac{1+\gamma_5}{2}
   +\frac{\partial^2W(\phi^*)}{\partial\phi_I^*\partial\phi_J^*}
   \frac{1-\gamma_5}{2}\right)
   \psi_J\biggr],
\label{eq:six}
\end{align}
where $D_{\text{w}}$ is the Wilson-Dirac operator,
\begin{equation}
   D_{\text{w}}\equiv
   \frac{1}{2}\sum_\mu\left\{\gamma_\mu(\partial_\mu^*+\partial_\mu)
   -a\partial_\mu^*\partial_\mu\right\}.
\label{eq:seven}
\end{equation}
In Ref.~\cite{Bergner:2007pu}, the authors consider
\begin{equation}
   \eta_I(\phi,\phi^*)=\frac{\partial W(\phi)}{\partial\phi_I}
   -\left(\partial_0^S-i\partial_1^S\right)\phi_I^*
   +i\frac{a}{2}\sum_\mu\partial_\mu^*\partial_\mu\phi_I.
\label{eq:eight}
\end{equation}
The resulting lattice action is given by~Eq.~\eqref{eq:six} with
$D_{\text{w}}\to\widetilde D_{\text{w}}$, where the ``twisted'' Wilson-Dirac
operator $\widetilde D_{\text{w}}$ is defined by
\begin{equation}
   \widetilde D_{\text{w}}\equiv
   \frac{1}{2}\sum_\mu\left\{\gamma_\mu(\partial_\mu^*+\partial_\mu)
   +ia\gamma_5\partial_\mu^*\partial_\mu\right\}.
\label{eq:nine}
\end{equation}
Finally, in Ref.~\cite{Kikukawa:2002as} (see also Ref.~\cite{Kawai:2010yj}),
\begin{align}
   &\eta_I(\phi,\phi^*)
\notag\\
   &=\frac{\partial W(\phi)}{\partial\phi_I}
   +\left(\phi_I^*-\frac{a}{2}\frac{\partial W(\phi^*)}{\partial\phi_I^*}\right)
   (S_0-iS_1)
   +\left(\phi_I-\frac{a}{2}\frac{\partial W(\phi)}{\partial\phi_I}\right)T,
\label{eq:ten}
\end{align}
where $S_\mu$ and $T$ denote the matrix elements of
\begin{align}
   &S_\mu=\frac{1}{2}(\partial_\mu^*+\partial_\mu)(A^\dagger A)^{-1/2},
\notag\\
   &T=\frac{1}{a}\left\{
   1-\left(1
   +\frac{1}{2}a^2\sum_\mu\partial_\mu^*\partial_\mu\right)(A^\dagger A)^{-1/2}
   \right\},
\label{eq:eleven}
\end{align}
and the combination $A\equiv 1-aD_{\text{w}}$ is defined from Wilson-Dirac
operator~\eqref{eq:seven}. The resulting lattice action is
\begin{align}
   &S_{\text{2DWZ}}^{\text{LAT}}
   =a^2\sum_x\biggl[
   -G_I^*G_I-G_I\eta_I(\phi,\phi^*)-G_I^*\eta_I^*(\phi,\phi^*)
\notag\\
   &\qquad
   +\overline\psi_I\left(D
   +\frac{1+\gamma_5}{2}
   \frac{\partial^2W(\phi)}{\partial\phi_I\partial\phi_J}
   \frac{1+\hat\gamma_5}{2}
   +\frac{1-\gamma_5}{2}
   \frac{\partial^2W(\phi^*)}{\partial\phi_I^*\partial\phi_J^*}
   \frac{1-\hat\gamma_5}{2}\right)
   \psi_J\biggr],
\label{eq:twelve}
\end{align}
where $D$ is the overlap-Dirac operator~\cite{Neuberger:1997fp,%
Neuberger:1998wv}
\begin{equation}
   D=\begin{pmatrix}T&S_0+iS_1\\S_0-iS_1&T\end{pmatrix},
\label{eq:thirteen}
\end{equation}
which fulfills the Ginsparg-Wilson relation~\cite{Ginsparg:1981bj},
\begin{equation}
   \gamma_5D+D\hat\gamma_5=0,\qquad\hat\gamma_5\equiv\gamma_5(1-aD).
\label{eq:fourteen}
\end{equation}
As a result of this relation~\cite{Luscher:1998pqa}, when the superpotential is
quasi-\hspace{0pt}homogeneous, the lattice action possesses an invariance under
the discrete subgroup $\mathbb{Z}_n$ of~$U(1)_A$~\cite{Kikukawa:2002as} (see
below).

\section{Perturbative proof of symmetry restoration in the
continuum limit}
The basic idea of the perturbative proof of symmetry restoration is common to
that of Refs.~\cite{Cohen:2003xe,Sugino:2003yb,Endres:2006ic,Kadoh:2009yf}:
assuming that symmetries under consideration do not suffer from the anomaly, in
the continuum limit, symmetry breaking owing to lattice regularization appears
only in local terms in the effective action, which correspond to 1PI diagrams
with non-negative superficial degree of divergence. Thus, we enumerate all
local (bosonic) polynomials of fields whose mass dimension is less than or
equal to two, because terms with the mass dimension higher than two correspond
to diagrams with negative superficial degree of divergence. The spacetime
integral of these local polynomials must be invariant under~$Q$,
Eq.~\eqref{eq:two}, and under~$U(1)_V$, Eq.~\eqref{eq:three}, because these are
manifest symmetries of the present lattice formulations. From mass dimensions
of fields and transformation law~\eqref{eq:two}, we see that the mass dimension
of~$Q$ is~$1/2$. Also, under~$U(1)_V$, Eq.~\eqref{eq:three}, $Q$ transforms as
\begin{equation}
   Q\to e^{i\alpha}Q,
\label{eq:fifteen}
\end{equation}
as again can be seen from transformation law~\eqref{eq:two}.

A key observation which allows for a quick enumeration of relevant local terms
is the triviality of the (local) $Q$-cohomology. From transformation
law~\eqref{eq:two}, it is easy to see that the $Q$-cohomology is trivial. That
is,
\begin{equation}
   QX([\varphi])=0\iff X([\varphi])=QY([\varphi])+\text{const.},
\label{eq:sixteen}
\end{equation}
where $[\varphi]$ collectively denotes all fields and $X$ and $Y$ are local
polynomials of fields at point $x$, for example. Moreover, by combining
Eq.~\eqref{eq:sixteen} with techniques of~Refs.~\cite{Brandt:1989rd,%
Brandt:1989gy,DuboisViolette:1991is,DuboisViolette:1992ye} (especially the
algebraic Poincar\'e lemma~\cite{Brandt:1989gy}), it is straightforward to show
that the local $Q$-cohomology is also trivial; this means,
\begin{equation}
   Q\int d^2x\,X([\varphi])=0\iff
   X([\varphi])=QY([\varphi])
   +\partial_\mu Z_\mu([\varphi])+\text{const.},
\label{eq:seventeen}
\end{equation}
where all $X$, $Y$ and $Z_\mu$ are local polynomials of fields. This shows that
in enumerating $Q$-invariant local terms in the effective action, we can
restrict ourselves to local polynomials of fields of the
form~$QY$. (Another possibility, a constant being independent of all fields,
has no physical consequence and can be neglected.) Here, the combination~$Y$
must contain an odd number of~$\psi_I$ (or~$\overline\psi_I$) for $QY$ to be
bosonic. Also, it should be proportional to at least one coupling constant
$\lambda_{\{m\}}^{(*)}$, because we are interested in terms induced by radiative
corrections (the classical continuum limit reproduces the target theory by
construction). Therefore, from the limitation of the mass dimension~$2$,
allowed local terms are at most linear in $\lambda_{\{m\}}^{(*)}$, $Q$
and~$\psi_I$ (or~$\overline\psi_I$). Taking also the $U(1)_V$ symmery into
account, only $\psi_I$ is possible. Thus, possible terms are given by a linear
combination of
\begin{align}
   &\lambda_{\{m\}}^{(*)}Q\left(f(\phi^*,\phi)\psi_{I\pm}\right)
\notag\\
   &=\lambda_{\{m\}}^{(*)}
   \left(-\frac{\partial f(\phi^*,\phi)}{\partial\phi_J^*}
   \overline\psi_{J+}\psi_{I\pm}
   -\frac{\partial f(\phi^*,\phi)}{\partial\phi_J}
   \overline\psi_{J-}\psi_{I\pm}
   +f(\phi^*,\phi)G_I^{(*)}\right),
\label{eq:eighteen}
\end{align}
where $f(\phi^*,\phi)$ is a local monomial of scalar fields. We can see,
however, that this combination cannot be induced by perturbative radiative
corrections in the above lattice formulations.

Let us first consider the lattice action, Eq.~\eqref{eq:six}
with~Eq.~\eqref{eq:five}. For example, the only way to have the last term
of~Eq.~\eqref{eq:eighteen} that is linear in~$\lambda_{\{m\}}^{(*)}$
and~$G_I^{(*)}$, is to connect scalar lines in the vertex
\begin{equation}
   a^2\sum_xG_I^{(*)}\frac{\partial W(\phi^{(*)})}{\partial\phi_I^{(*)}},
\label{eq:nineteen}
\end{equation}
to make a 1PI tadpole diagram. However, to have such a diagram, we need a free
propagator between $\phi_J$ and~$\phi_K$, $\langle\phi_J(x)\phi_K(y)\rangle_0$
(or between $\phi_J^*$ and~$\phi_K^*$,
$\langle\phi_J^*(x)\phi_K^*(y)\rangle_0$). As can easily be verified from
Eqs.~\eqref{eq:six} and~\eqref{eq:five}, free propagators of these types
identically vanish. Note that the lattice action possesses the invariance under
$\phi_I\to e^{-i\alpha}\phi_I$ and~$\phi_I^*\to e^{i\alpha}\phi_I^*$ in the free
theory. Thus the last term of~Eq.~\eqref{eq:eighteen} cannot be induced by
radiative corrections.

The situation is similar for other terms in~Eq.~\eqref{eq:eighteen}. To have
the term containing $\overline\psi_{J-}\psi_{I+}$, for example, we have to
connect scalar lines in the Yukawa interaction
\begin{equation}
   a^2\sum_x\overline\psi_J
   \frac{\partial^2W(\phi)}{\partial\phi_J\partial\phi_I}
   \frac{1+\gamma_5}{2}\psi_I,
\label{eq:twenty}
\end{equation}
to make a tadpole. This is again impossible, because we do not have a free
propagator of the type $\langle\phi_K(x)\phi_L(y)\rangle_0$.

From these considerations, we observe that \emph{no\/} local term that
corresponds to a 1PI diagram with non-negative superficial degree of divergence
is induced by perturbative radiative corrections to the effective action. From
this, we infer that all symmetries broken by the lattice regularization are
restored in the continuum limit to all orders in perturbation
theory.\footnote{In this regard, one of us (H.S.) would like to apologize the
authors of~Ref.~\cite{Sakai:1983dg} for his wrong statement made
in~Ref.~\cite{Suzuki:2010} that a discrete lattice symmetry~$\mathbb{Z}_n$ (see
below), which the lattice formulation of~Ref.~\cite{Sakai:1983dg} does not
have, is crucial for the SUSY restoration. In reality, as shown above, the
discrete lattice symmetry is not indispensable for the SUSY restoration.}
Note that the fact that $\lambda_{\{m\}}^{(*)}$ are dimensionful and the present
2D system is super-renormalizable is crucial for the above proof.

The argument goes almost identically for other lattice actions, because they
have common features: $Q$ and~$U(1)_V$ invariance\footnote{One can easily
modify the above proof so that it does not require the $U(1)_V$ invariance.}
and no free propagators of the types~$\langle\phi_I(x)\phi_J(y)\rangle_0$
and~$\langle\phi_I^*(x)\phi_J^*(y)\rangle_0$, as can easily be verified.

\section{Conclusion}
In this article, we have shown to all orders in perturbation theory that for
lattice formulations of 2D $\mathcal{N}=(2,2)$ WZ model on the basis of the
lattice Nicolai map, Eqs.~\eqref{eq:five}, \eqref{eq:eight} and~\eqref{eq:ten},
SUSY and other symmetries broken by lattice regularization are restored in the
continuum limit without fine tuning. Our this result provides a theoretical
basis for using these lattice formulations for computation of correlation
functions.

All the above lattice formulations are thus equivalent in the sense that they
all require no fine tuning to reach a SUSY point in the continuum limit. The
way of approaching the continuum theory can, however, be different. Generally
speaking, a lattice formulation might be regarded superior if higher
symmetries are preserved with it. In this respect, the formulation
with~Eq.~\eqref{eq:ten} is superior, because it possesses a higher symmetry
when the superpotential is quasi-homogeneous~\cite{Kikukawa:2002as}. When the
superpotential is quasi-homogeneous (see footnote~\ref{fn:one}),
\begin{equation}
   W(\phi_I\to e^{i\omega_I\alpha}\phi_I)=e^{i\alpha}W(\phi),
\label{eq:twentyone}
\end{equation}
and thus the continuum action (after integrating over the auxiliary fields)
possesses an invariance under a $U(1)_A$ transformation that is given by,
\begin{align}
   &\phi_I\to e^{i\omega_I\alpha}\phi_I,
   &&\phi_I^*\to e^{-i\omega_I\alpha}\phi_I^*,
\notag\\
   &\psi_I\to e^{i(\omega_I-1/2)\alpha\gamma_5}\psi_I,&&
   \overline\psi_I\to\overline\psi_Ie^{i(\omega_I-1/2)\alpha\gamma_5}.
\label{eq:twentytwo}
\end{align}
This symmetry cannot be promoted to a lattice symmetry in the cases of
Eq.~\eqref{eq:five} and~Eq.~\eqref{eq:eight}, because the resulting (twisted)
Wilson term cannot be compatible with the chiral $\gamma_5$~rotation. With the
choice~\eqref{eq:ten}, on the other hand, thanks to the Ginsparg--Wilson
relation~\eqref{eq:fourteen}, the part of the action quadratic in the spinor
field possesses a lattice $U(1)_A$ symmetry corresponding
to~Eq.~\eqref{eq:twentytwo}:
\begin{align}
   &\phi_I\to e^{i\omega_I\alpha}\phi_I,
   &&\phi_I^*\to e^{-i\omega_I\alpha}\phi_I^*,
\notag\\
   &\psi_I\to e^{i(\omega_I-1/2)\alpha\hat\gamma_5}\psi_I,&&
   \overline\psi_I\to\overline\psi_Ie^{i(\omega_I-1/2)\alpha\gamma_5}.
\label{eq:twentythree}
\end{align}
Although this $U(1)_A$ invariance for arbitrary~$\alpha$ is broken by a
term in the lattice action (after integrating over the auxiliary fields),
\begin{equation}
   -\frac{\partial W(\phi)}{\partial\phi_I}(S_0+iS_1)\phi_I,
\label{twentyfour}
\end{equation}
the so-called would-be surface term~\cite{Kikukawa:2002as} (and its complex
conjugate), not all is lost. Since the above would-be surface term is also
quasi-homogeneous with same weights~$\omega_I$ as~$W(\phi)$, a discrete
subgroup $\mathbb{Z}_n$ of~$U(1)_A$, which is given
by~Eq.~\eqref{eq:twentythree} with the angles $\alpha=2\pi k$, $k=0$, $1$,
$2$, \dots, $n-1$ (where the integer $n$ is determined by weights~$\omega_I$),
remains an exact lattice symmetry. This exact lattice symmetry could imply a
faster approach to the continuum theory; this point deserves further study.

We are indebted to Yoshio Kikukawa for helpful discussions in the initial
stage of this work. We would like to thank Michael G.~Endres for a careful
reading of the manuscript and Fumihiko Sugino for an informative discussion.
One of us (H.S.) would like to thank Ting-Wai Chiu for the hospitality extended
to him at the National Taiwan University, where this work was completed. The
work of H.S.\ is supported in part by a Grant-in-Aid for Scientific Research,
22340069.





\bibliographystyle{elsarticle-num}
\bibliography{<your-bib-database>}

\begin{thebibliography}{00}


\bibitem{Kastor:1988ef}
  D.~A.~Kastor, E.~J.~Martinec and S.~H.~Shenker,
  Nucl.\ Phys.\  B {\bf 316} (1989) 590.

\bibitem{Vafa:1988uu}
  C.~Vafa and N.~P.~Warner,
  Phys.\ Lett.\  B {\bf 218} (1989) 51.

\bibitem{Lerche:1989uy}
  W.~Lerche, C.~Vafa and N.~P.~Warner,
  Nucl.\ Phys.\  B {\bf 324} (1989) 427.

\bibitem{Howe:1989qr}
  P.~S.~Howe and P.~C.~West,
  Phys.\ Lett.\  B {\bf 223} (1989) 377.

\bibitem{Cecotti:1989jc}
  S.~Cecotti, L.~Girardello and A.~Pasquinucci,
  Nucl.\ Phys.\  B {\bf 328} (1989) 701.

\bibitem{Howe:1989az}
  P.~S.~Howe and P.~C.~West,
  Phys.\ Lett.\  B {\bf 227} (1989) 397.

\bibitem{Cecotti:1989gv}
  S.~Cecotti, L.~Girardello and A.~Pasquinucci,
  Int.\ J.\ Mod.\ Phys.\  A {\bf 6} (1991) 2427.

\bibitem{Cecotti:1990kz}
  S.~Cecotti,
  Int.\ J.\ Mod.\ Phys.\  A {\bf 6} (1991) 1749.

\bibitem{Witten:1993jg}
  E.~Witten,
  Int.\ J.\ Mod.\ Phys.\  A {\bf 9} (1994) 4783
  [arXiv:hep-th/9304026].

\bibitem{Hori:2003ic}
  K.~Hori {\it et al.},
  ``Mirror symmetry,''
{\it  Providence, USA: AMS (2003) 929~p}

\bibitem{Kawai:2010yj}
  H.~Kawai and Y.~Kikukawa,
  arXiv:1005.4671 [hep-lat].

\bibitem{Kikukawa:2002as}
  Y.~Kikukawa and Y.~Nakayama,
  Phys.\ Rev.\  D {\bf 66} (2002) 094508
  [arXiv:hep-lat/0207013].

\bibitem{Beccaria:1998vi}
  M.~Beccaria, G.~Curci and E.~D'Ambrosio,
  Phys.\ Rev.\  D {\bf 58} (1998) 065009
  [arXiv:hep-lat/9804010].

\bibitem{Catterall:2001fr}
  S.~Catterall and S.~Karamov,
  Phys.\ Rev.\  D {\bf 65} (2002) 094501
  [arXiv:hep-lat/0108024].

\bibitem{Giedt:2005ae}
  J.~Giedt,
  Nucl.\ Phys.\  B {\bf 726} (2005) 210
  [arXiv:hep-lat/0507016].

\bibitem{Bergner:2007pu}
  G.~Bergner, T.~Kaestner, S.~Uhlmann and A.~Wipf,
  Annals Phys.\  {\bf 323} (2008) 946
  [arXiv:0705.2212 [hep-lat]].

\bibitem{Kastner:2008zc}
  T.~K\"astner, G.~Bergner, S.~Uhlmann, A.~Wipf and C.~Wozar,
  Phys.\ Rev.\  D {\bf 78} (2008) 095001
  [arXiv:0807.1905 [hep-lat]].

\bibitem{Kanamori:2008bk}
  I.~Kanamori and H.~Suzuki,
  Nucl.\ Phys.\  B {\bf 811} (2009) 420
  [arXiv:0809.2856 [hep-lat]].

\bibitem{Giedt:2004qs}
  J.~Giedt and E.~Poppitz,
  JHEP {\bf 0409} (2004) 029
  [arXiv:hep-th/0407135].

\bibitem{Sakai:1983dg}
  N.~Sakai and M.~Sakamoto,
  Nucl.\ Phys.\  B {\bf 229} (1983) 173.

\bibitem{Nicolai:1979nr}
  H.~Nicolai,
  Phys.\ Lett.\  B {\bf 89} (1980) 341.

\bibitem{Nicolai:1980jc}
  H.~Nicolai,
  Nucl.\ Phys.\  B {\bf 176} (1980) 419.

\bibitem{Cecotti:1981fu}
  S.~Cecotti and L.~Girardello,
  Phys.\ Lett.\  B {\bf 110} (1982) 39.

\bibitem{Parisi:1982ud}
  G.~Parisi and N.~Sourlas,
  Nucl.\ Phys.\  B {\bf 206} (1982) 321.

\bibitem{Cecotti:1982ad}
  S.~Cecotti and L.~Girardello,
  Nucl.\ Phys.\  B {\bf 226} (1983) 417.

\bibitem{Bartels:1983wm}
  J.~Bartels and J.~B.~Bronzan,
  Phys.\ Rev.\  D {\bf 28} (1983) 818.

\bibitem{Kadoh:2009sp}
  D.~Kadoh and H.~Suzuki,
  Phys.\ Lett.\  B {\bf 684} (2010) 167
  [arXiv:0909.3686 [hep-th]].

\bibitem{Sugino:2008yp}
  F.~Sugino,
  Nucl.\ Phys.\  B {\bf 808} (2009) 292
  [arXiv:0807.2683 [hep-lat]].

\bibitem{Elitzur:1982vh}
  S.~Elitzur, E.~Rabinovici and A.~Schwimmer,
  Phys.\ Lett.\  B {\bf 119} (1982) 165.

\bibitem{Elitzur:1983nj}
  S.~Elitzur and A.~Schwimmer,
  Nucl.\ Phys.\  B {\bf 226} (1983) 109.

\bibitem{Neuberger:1997fp}
  H.~Neuberger,
  Phys.\ Lett.\  B {\bf 417} (1998) 141
  [arXiv:hep-lat/9707022].

\bibitem{Neuberger:1998wv}
  H.~Neuberger,
  Phys.\ Lett.\  B {\bf 427} (1998) 353
  [arXiv:hep-lat/9801031].

\bibitem{Ginsparg:1981bj}
  P.~H.~Ginsparg and K.~G.~Wilson,
  Phys.\ Rev.\  D {\bf 25} (1982) 2649.

\bibitem{Luscher:1998pqa}
  M.~L\"uscher,
  Phys.\ Lett.\  B {\bf 428} (1998) 342
  [arXiv:hep-lat/9802011].

\bibitem{Cohen:2003xe}
  A.~G.~Cohen, D.~B.~Kaplan, E.~Katz and M.~\"Unsal,
  JHEP {\bf 0308} (2003) 024
  [arXiv:hep-lat/0302017].

\bibitem{Sugino:2003yb}
  F.~Sugino,
  JHEP {\bf 0401} (2004) 015
  [arXiv:hep-lat/0311021].

\bibitem{Endres:2006ic}
  M.~G.~Endres and D.~B.~Kaplan,
  JHEP {\bf 0610} (2006) 076
  [arXiv:hep-lat/0604012].

\bibitem{Kadoh:2009yf}
  D.~Kadoh, F.~Sugino and H.~Suzuki,
  Nucl.\ Phys.\  B {\bf 820} (2009) 99
  [arXiv:0903.5398 [hep-lat]].

\bibitem{Brandt:1989rd}
  F.~Brandt, N.~Dragon and M.~Kreuzer,
  Phys.\ Lett.\  B {\bf 231} (1989) 263.

\bibitem{Brandt:1989gy}
  F.~Brandt, N.~Dragon and M.~Kreuzer,
  Nucl.\ Phys.\  B {\bf 332} (1990) 224.

\bibitem{DuboisViolette:1991is}
  M.~Dubois-Violette, M.~Henneaux, M.~Talon and C.~M.~Viallet,
  Phys.\ Lett.\  B {\bf 267} (1991) 81.

\bibitem{DuboisViolette:1992ye}
  M.~Dubois-Violette, M.~Henneaux, M.~Talon and C.~M.~Viallet,
  Phys.\ Lett.\  B {\bf 289} (1992) 361
  [arXiv:hep-th/9206106].

\bibitem{Suzuki:2010}
H.~Suzuki,
``Lattice SUSY from a practitioner's perspective'',
a talk given at the YITP workshop YITP-W-10-11 on
``Discretization approaches to the dynamics of space-time and fields''
(27 September, 2010).

\end{thebibliography}



\end{document}